\documentclass[article,osajnl,showpacs,superscriptaddress,twocolumn,10.5pt]{revtex4-2}

\usepackage{graphicx}
\usepackage{dcolumn}
\usepackage{bm}
\usepackage{subfigure}
\usepackage{blindtext}
\usepackage{float}
\usepackage{multirow}
\makeatletter
\newcommand{\printfnsymbol}[1]{%
  \textsuperscript{\@fnsymbol{#1}}%
}
\makeatother

\begin{document}

\title{EEG functional connectivity and deep learning for automatic diagnosis of brain disorders: Alzheimer’s disease and schizophrenia}

\affiliation{BioMEMS, Technische Hochschule Aschaffenburg, Aschaffenburg, Germany}
\affiliation{Universidade de São Paulo (USP), \\ Instituto de Ciências Matemáticas e de Computação,  São Carlos,  SP,  Brazil}

\author{Caroline L. Alves\printfnsymbol{1} }
\email{caroline.alves@th-ab.de}
\affiliation{BioMEMS, Technische Hochschule Aschaffenburg, Aschaffenburg, Germany}
\affiliation{Universidade de São Paulo (USP), \\ Instituto de Ciências Matemáticas e de Computação,  São Carlos,  SP,  Brazil}

\author{Aruane M. Pineda\thanks{equal contribution}}
\email{aruane.pineda@usp.br}
\affiliation{Universidade de São Paulo (USP), \\ Instituto de Ciências Matemáticas e de Computação,  São Carlos,  SP,  Brazil}

\author{Kirstin Roster}
\affiliation{Universidade de São Paulo (USP), \\ Instituto de Ciências Matemáticas e de Computação,  São Carlos,  SP,  Brazil}
 
\author{Christiane Thielemann}
\affiliation{BioMEMS, Technische Hochschule Aschaffenburg, Aschaffenburg, Germany}
 
\author{Francisco A. Rodrigues}
\affiliation{Universidade de São Paulo (USP), \\ Instituto de Ciências Matemáticas e de Computação,  São Carlos,  SP,  Brazil}
 
\begin{abstract}
Mental disorders are among the leading causes of disability worldwide. The first step in treating these conditions is to obtain an accurate diagnosis, but the absence of established clinical tests makes this task challenging. Machine learning algorithms can provide a possible solution to this problem, as we describe in this work. We present a method for the automatic diagnosis of mental disorders based on the matrix of connections obtained from EEG time series and deep learning. We show that our approach can classify patients with Alzheimer’s disease and schizophrenia with a high level of accuracy. The comparison with the traditional cases, that use raw EEG time series, shows that our method provides the highest precision.  Therefore, the application of deep neural networks on data from brain connections is a very promising method to the diagnosis of neurological disorders.
\end{abstract}

\maketitle
\section{Introduction}

Neurological disorders, including Alzheimer's disease (AD) and schizophrenia (SZ),  are among the main priorities in the present global health agenda \cite{world2006neurological}. AD is a type of dementia that affects primarily elderly individuals and is characterized by the degeneration of brain tissue, leading to impaired intellectual and social abilities \cite{dolgin2016defeat}. Currently, around 25 million people live with AD \cite{Organization/2021}. In the US, nearly six million individuals are affected by AD, with incidence projected to increase more than two-fold to 13.8 million by 2050 \cite{rodriguez2021machine}. Individuals with SZ have symptoms such as hallucinations, incoherent thinking, delusions, decreased intellectual functioning, difficulty in expressing emotions, and agitation \cite{jahmunah2019automated,gottesman1982schizophrenia}. According to the World Health Organization (WHO), SZ affects around 26 million people worldwide \cite{Organization/20212}. 

The base for a successful treatment of AD and SZ is the correct diagnosis. However, in the absence of established clinical tests for neurological disorders, both the diagnosis and the determination of the stage of AD and SZ are based primarily on qualitative interviews, including psychiatric history and current symptoms, and the assessment of behaviour. These observations may be subjective, imprecise, and incomplete \cite{borsboom2013network,Fried2017mental,borsboom2008psychometric}. 
To provide a quantitative evaluation of mental disorders, methods based on Magnetic Resonance Imaging (MRI), Computerized Tomography (CT) \cite{jack2009serial}, and Positron Emission Tomography (PET) \cite{ding2019deep,walhovd2010combining} has been used to aid professionals in the diagnostic process  \cite{del2018trimage}. However, the use of multiple imaging devices can be expensive to implement and the fusion of images from different devices can have poor quality due to motion artifacts.

\begin{figure*}[t]
\centering
\includegraphics[width=0.99\textwidth]{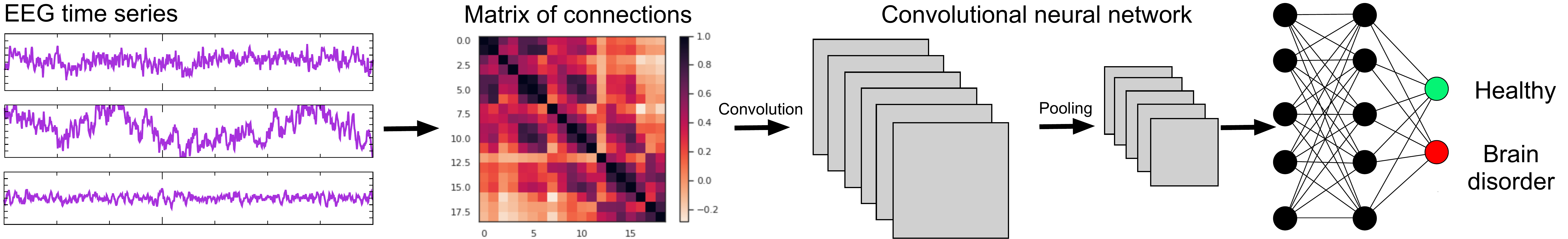}
\caption{Illustration of the method for automatic diagnosis of mental disorders based on EEG time series. Time series are collected and the correlation between electrodes are calculated yielding the matrices of connections, which encompass the functional connectivity between brain regions. Finally, the CNN is adjusted to enable the automatic classification of individuals.}
\label{Fig:methodology}
\end{figure*}

To overcome these restrictions, EEG data is a viable candidate to support the diagnosis of SZ and AD  \cite{tait2020eeg}. Although EEG has a low spatial resolution, it has a comparatively low cost, good temporal resolution and is easily available in most contexts. Nonetheless, visual analysis of EEG data is time-consuming, requires specialized training, and is error-prone \cite{trambaiolli2011eeg,falk2012eeg,piubelli2021serum}. However,  we can consider automatic evaluation of EEG time series using modern classification algorithms, which can help to improve the efficiency and accuracy of AD and SZ diagnosis, as verified in previous works \cite{pineda2020quantile,oh2019deep,ahmadlou2011fractality,buettner2020development}. 

Moreover, instead of using raw EEG time series, it is possible to encompass the connections between brain regions by constructing cortical complex networks~\cite{sporns2002network}. In this case, we build cortical networks for healthy and individuals with neurological disorders. To distinguish between them, we  use network measures to describe the network structure, as described in a previous work of ours \cite{de2014structure} (see also \cite{costa2007characterization, da2010pattern} for a description of the methodology used in network classification). Therefore, each network is mapped into a $d$ dimensional space, where $d$ is the number of measures adopted for network characterization. This process of building a set of features to represent the input data is called feature engineering. After extracting the network features for the two classes of networks, i.e. healthy and individuals with mental disorders, supervised learning algorithms are adjusted to perform automatic classification. Previous works verified that this approach enables the diagnosis with accuracy higher than $80\%$ in the case of childhood-onset schizophrenia~\cite{de2014structure}. 

Although this methodology has been used for many different diseases (e.g. \cite{pineda2020quantile,de2014structure,diykh2017classify}) the performance of the algorithm depends on the measures selected to describe the network structure. The network properties included in the model could represent just a subset of the information necessary to get the best performance of the supervised model. Therefore, the network representation can be incomplete, restricting the accuracy of the classifiers. One possible solution to this problem is the use of a matrix of connections in combination with  and deep neural networks~\cite{goodfellow2016deep}, as we show in the present paper. In this case, instead of extracting the network measures, the matrix of connections is considered as input to train a deep neural network. This matrix encodes all the information necessary to represent the network structure and avoid the choice of network measures.

Therefore, we consider  the matrix of connections between brain areas and deep neural networks to distinguish individuals with AD and SZ from healthy controls. Other than previous works, where only raw time series are adopted as input for the neural network\cite{acharya2018deep,kashiparekh2019convtimenet,islam2018brain,duneja2019analysis,acharya2018automated,oh2018deep,yildirim2018deep}, we do not ignore the connections between the electrodes used to record the time series. We construct the matrix of connections by using Granger causality, Pearson's and Spearman's correlations \cite{bastos2016tutorial,seth2015granger,bonita2014time}. We verify that this information about the connections is fundamental and improves the classification, compared to the previously mentioned approaches that use only raw EEG time series. 

In summary, in this work we achieve the following contributions:
\begin{itemize}
\item We propose a method to classify EEG time series from healthy and patients presenting AD and SZ.  With a matrix of connections as input for  a tuned Convolutional Neural Network (CNN) model, the accuracy obtained is close to 100 \% for both disorders. Our results are more accurate than those observed in previous works that consider only raw EEG time series, reinforcing the importance of the network structure on the diagnosis of mental disorders.

\item We show that the method to infer the matrices of connections influences the quality of the classification results. For SZ, the Granger causality provides the most accurate classification, whereas, for AD, the Pearson's correlation yields the highest accuracy.

\item Our framework is general and can be used in EEG data from any brain disorder. It allows to determine the best cortical network representation and adjust the CNN to optimize the accuracy.

\end{itemize}

In the next sections, we outline the data set, present the CNN architecture and show our results, comparing them with more common approaches that do not consider the connections between brain areas. 

\section{EEG data} \label{Sec:database}

The AD data set considered here is composed of EEG time series recorded at a sampling frequency of 128~Hz and a duration of eight seconds for each individual and at 19 channels ($F_{p1}$, $F_{p2}$, $F_{7}$, $F_{3}$, $F_{z}$, $F_{4}$, $F_{8}$, $T_{3}$, $C_{3}$, $C_{z}$, $C_{4}$, $T_{4}$, $T_{5}$, $P_{3}$, $P_{z}$, $P_{4}$, $T_{6}$, $O_{1}$, and $O_{2}$)~\cite{pineda2020quantile,pritchard1991altered}. The letters F, C, P, O, and T refer to the respective cerebral lobes frontal (F), central (C), parietal (P), occipital (O), and temporal (T). The data is divided into two sets. The first one consists of 24 healthy elderly individuals (control group; aged 72 ± 11 years) who do not have any history of neurological disorders. The second one is made of 24 elderly individuals with AD (aged 69 ± 16 years) diagnosed by the National Institute of Neurological and Communicative Disorders and Stroke, the Alzheimer’s Disease and Related Disorders Association (NINCDS-ADRDA), following the Diagnostic and Statistical Manual of Mental Disorders (DSM)-III-R criteria (\cite{pineda2020quantile,pritchard1991altered}).

The data set used for diagnosis of SZ can be found at \cite{timashev2012analysis}. This data contains 16-channel EEG time series recorded at a sampling frequency of \nolinebreak{128 Hz} over one minute, including $F_{7}$, $F_{3}$, $F_{4}$, $F_{8}$, $T_{3}$, $C_{3}$, $C_{z}$, $C_{4}$, $T_{4}$, $T_{5}$, $P_{3}$, $P_{z}$, $P_{4}$, $T_{6}$, $O_{1}$, and $O_{2}$. Notice that both data set come from studies of 16 common brain regions, with the AD data set having three more regions analyzed. Furthermore, it also includes two sets, (i) one of 39 healthy young individuals (control group; aged 11 to 14 years) and (ii) one of 45 teenagers individuals (aged 11 to 14 years) with symptoms of schizophrenia.

\section{Concepts and Methods}
\label{Sec:concepts}

Our framework to perform the automatic diagnosis of AD and SZ is illustrated in Figure \ref{Fig:methodology}. In a first step, EEG time series, which are free of artifacts, are used to construct the matrices of connections.  The strength of the connections between two brain regions is quantified by three different methods: (i) Granger causality test \cite{granger1969investigating}, (ii) the Pearson's  ~\cite{benesty2009pearson} and (iii) Spearman's~\cite{lubinski2004introduction} correlation measures. For Granger causality, a statistical hypothesis test is done and p-values are obtained. The matrices are filled with ``1'' if $p < 0.05$ and ``0'' if $p\geq 0.05$. Matrices are calculated for AD data sets (19 EEG channels) and for SZ data sets (16 EEG channels). These matrices are inserted in a CNN to discriminate healthy individuals from individuals diagnosed with AD and SZ (see Figure \ref{Fig:methodology}).
Notice that the use of different methods to infer the brain areas is necessary because there is no general method to infer functional connectivity~\cite{bastos2016tutorial,seth2015granger,bonita2014time, comin2020complex}. 
Indeed, choosing the best metric to infer these connections between brain areas is a current challenge in network neuroscience (e.g.~\cite{shandilya2011inferring, lusch2016inferring}).

\subsection{Convolutional Neural Network}
\label{Sec:CNN}

CNN is a type of neural network \cite{millstein2020convolutional} with three types of layers and masked parameters, as proposed in \cite{hubel1962receptive, lopez2002convolutional}. 
The convolutional layer performs the mathematical operation called convolution, which is done in more than one dimension at a time. The weights of the artificial neurons are represented by a tensor called kernel (or filter). The outputs from the convolutional layer include the main features from the input data. The convolution process between neurons and kernels produces outputs called feature maps.

The pooling layer reduces the dimensionality and operates similarly to the convolutional layer. The difference is that pooling kernels are weightless and add aggregation functions to their input data, such as a maximum or mean function \cite{lecun1989generalization,lecun2015deep}. The max pooling function is used here to return the highest value within an area of the tensor, which reduces the size of the feature map.
The fully connected layer categorizes input data into different classes, based on an initial set of data used for training. The artificial neurons in the max pooling and fully connected layers are connected, as the output predicts precisely the result of the input EEG data as healthy and unhealthy \cite{oh2019deep}.

Two approaches for the CNN architectures are proposed here, one using a tuning method (CNN$_{tuned}$) and another without this optimization step (CNN$_{untuned}$).  Tuning is an optimization method used to find the values of hyperparameters to improve the performance of the CNN model \cite{hutter2015beyond}. Three tuning techniques are used in the present work: (i) random search \cite{bergstra2012random},  (ii) hyper-band \cite{rostamizadeh2017efficient} and (iii) Bayesian optimization  \cite{doke2020using}. The traditional way to optimize the hyperparameters is exhaustive searching through a manually specified parameters search space and evaluate all possible combinations of these parameter values. However, this approach has a high computational cost.  An alternative method is to select the values of parameters in the search space at random until maximize the objective function (here, this objective function is the maximization of accuracy). 

The idea of hyper-band optimizations is to select different possible models (with different hyperparameters values), train them for a time, and discard the worst one at each iteration, until a few combinations remain.  
In contrast, Bayesian optimization is a global optimization method that uses the Bayes Theorem to direct the search to find the minimum or maximum of a certain objective function  \cite{goodfellow2016deep}.  

In the CNN$_{tuned}$ model, the dropout regularization technique is employed to avoid overfitting~\cite{srivastava2014dropout}. The layers and range used for hyperparameters are presented in table \ref{Tab:hyper}. The best CNN$_{tuned}$ architectures tuned for each data set individually are depicted in table \ref{table:tuning-sz-ad}.  The CNN$_{untuned}$ model presents fewer layers and therefore lower computational costs. The parameters used in our analysis are described in table \ref{Table:cnn-untuned}. 

\begin{table*}[!t]
\centering
\caption{Best hyperparameters and layer configurations obtained for the  CNN$_{tuned}$ model.}
\vspace{3mm}
\begin{tabular}{cccc}
			\hline
			\textbf{Type of Layer}                & \textbf{Tuning hyperparameter} &           \textbf{Value}         \\  
			\hline
			
			Convolutional      &---   &---  \\
			\hline
			                              & & [0.00, 0.05, 0.10, 0.15, \\
			Convolutional        &dropout  &  0.20, 0.25, 0.30,   \\
			        & & 0.35, 0.40, 0.45, 0.50]  \\
			\hline
			Convolutional       & ---  & ---  \\
			\hline
				\hline
			Convolutional       & number of filters  & [32, 64] \\
			\hline
				\hline
			Max Pooling       & dropout  & [0.00, 0.50, 0.10, 0.15, 0.20]  \\
			\hline
			Flatten     & ---  & ---  \\
			\hline
			Dense&            - units&[32, 64, 96....512]\\
			       & -activation  & [relu, tanh, sigmoid]\\
			\hline
			Dropout      & rate  & [0.00, 0.50, 0.10, 0.15, 0.20]  \\
			\hline
				Adam  &    & $min-value =1e^{-4}$ \\
			optimization         &  learning  & $max-value=1e^{-2}$     \\
			compile          &  rate & sampling= LOG   \\
			\hline
			
		\end{tabular}
\label{Tab:hyper}
\end{table*}	

\begin{table*} [!t]
\centering
\caption{The network architecture for the CNN$_{tuned}$ model used in the AD and SZ data sets.}
\vspace{3mm}
\begin{tabular}{ccccc} 
			\hline
			\textbf{Type of Layer} & \textbf{Output Shape (AD)} &
			\textbf{Output Shape (SZ)}
			&\textbf{Parameter}\\  
			\hline
			Convolutional &(None, 17, 17, 16) &(None, 14, 14, 16)    &160  \\
			\hline
			Convolutional    &(None, 15, 15, 16) &(None, 12, 12, 16) &2320\\
			\hline
			max-pooling        &(None, 7, 7, 16) &(None, 6, 6, 16)&0 \\
			\hline
			dropout         & (None, 7, 7, 16)&(None, 6, 6, 16)&0 \\
			\hline
			Convolutional       &(None, 5, 5, 32)&(None, 4, 4, 32) &4640 \\
			\hline
			Convolutional        & (None, 3, 3, 32)& (None, 2, 2, 32)&9248 \\
			\hline
			max-pooling    & (None, 1, 1, 32)& (None, 1, 1, 32)&0 \\
			\hline
			dropout       &(None, 1, 1, 32) & (None, 1, 1, 32)&0  \\
			\hline
			flatten       & (None, 32) & (None, 32)&0\\
			\hline
			dense      & (None, 160) & (None, 160) &5280  \\
			\hline
			dropout       &(None, 160)  & (None, 160) &0 \\
	        \hline
	        dense        &(None, 2)  & (None, 2)&3 \\
			\hline
		\end{tabular}
\label{table:tuning-sz-ad}
\end{table*}

\begin{table*}
\centering
\caption{The network architecture for the CNN$_{untuned}$ model used in the AD and SZ data sets.}
\begin{tabular}{ccccc}
\hline
\textbf{Type of Layer}                       & \textbf{Output Layer (AD)} &  \textbf{Output Layer (SZ)} &\textbf{Kernel} \\ \hline
 \textbf{Input Layer}                         & 19 x 19 x 1                            & 16x16x1 &-      \\        \hline
\textbf{Convolution}                         & 18 x 18 x 32 &15 x 15 x 32&    4       \\
\hline
\textbf{Max pooling}                         & 18 x 18 x 32&15 x 15 x 32                           & 2         \\
\hline
\textbf{Convolution}                         & 17 x17 x 16 & 14 x 14 x 16                            & 4      \\
\hline
 \textbf{Max pooling}                         & 17 x17 x 16   & 14 x 14 x16                         & 2        \\
\hline
 \textbf{Flatten}                             & 17 x17 x 16 & 3136                            & -               \\
\hline
\multicolumn{1}{l}{\textbf{Fully connected}} & 10 &10                                    & -               \\
\hline
\textbf{Fully connected}                     & 1    &1                                  &-               \\ \hline
\end{tabular}
\label{Table:cnn-untuned}
\end{table*}

\subsection{Evaluation}

Since we have a two-class classification problem (negative and positive), we consider the precision and recall measures in the evaluation process \cite{maimon2010data}. Precision (also called specificity) corresponds to the hit rate in the negative class, whereas recall (also called sensitivity)  measures how well a classifier can predict positive examples (hit rate in the positive class).
For visualization of these two measures, the receiver operating characteristic (ROC) curve is a common method as it displays the relation between the rate of true positives and false positives. The area below this curve, called area under ROC curve (AUC) has been widely used in classification problems \cite{huang2005using}, mainly for medical diagnoses \cite{ozcift2011classifier,shen2016evolving,tanwani2009guidelines,nanni2010local}. The value of the AUC varies from 0 to 1, where the value of one corresponds to a classification result free of errors. $AUC = 0.5$ indicates that the classifier is not able to distinguish the two classes --- this result is equal to the random choice. Furthermore, we consider the micro average of ROC curve, which computes the AUC metric independently for each class (calculate AUC metric for healthy individuals, class zero, and separately calculate for unhealthy subjects, class one) and then the average is computed considering these classes equally. The macro average is also used in our evaluation, which does not consider both classes equally, but aggregates the contributions of the classes separately and then calculates the average.

\section{Results and discussion}

\begin{figure*}\centering
\subfigure[]{
{\includegraphics[width=0.450\textwidth]{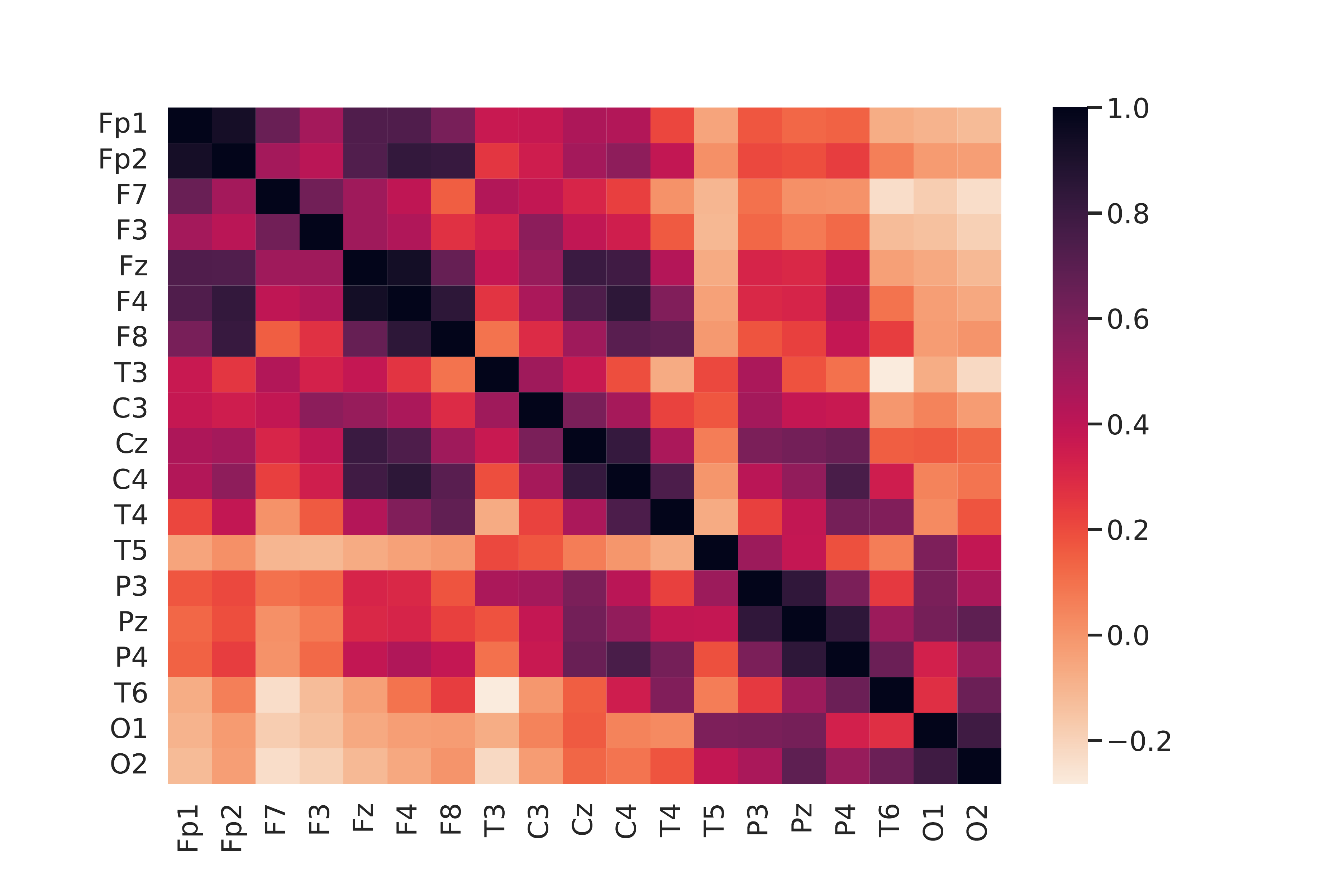}}}
\subfigure[]{
{\includegraphics[width=0.450\textwidth]{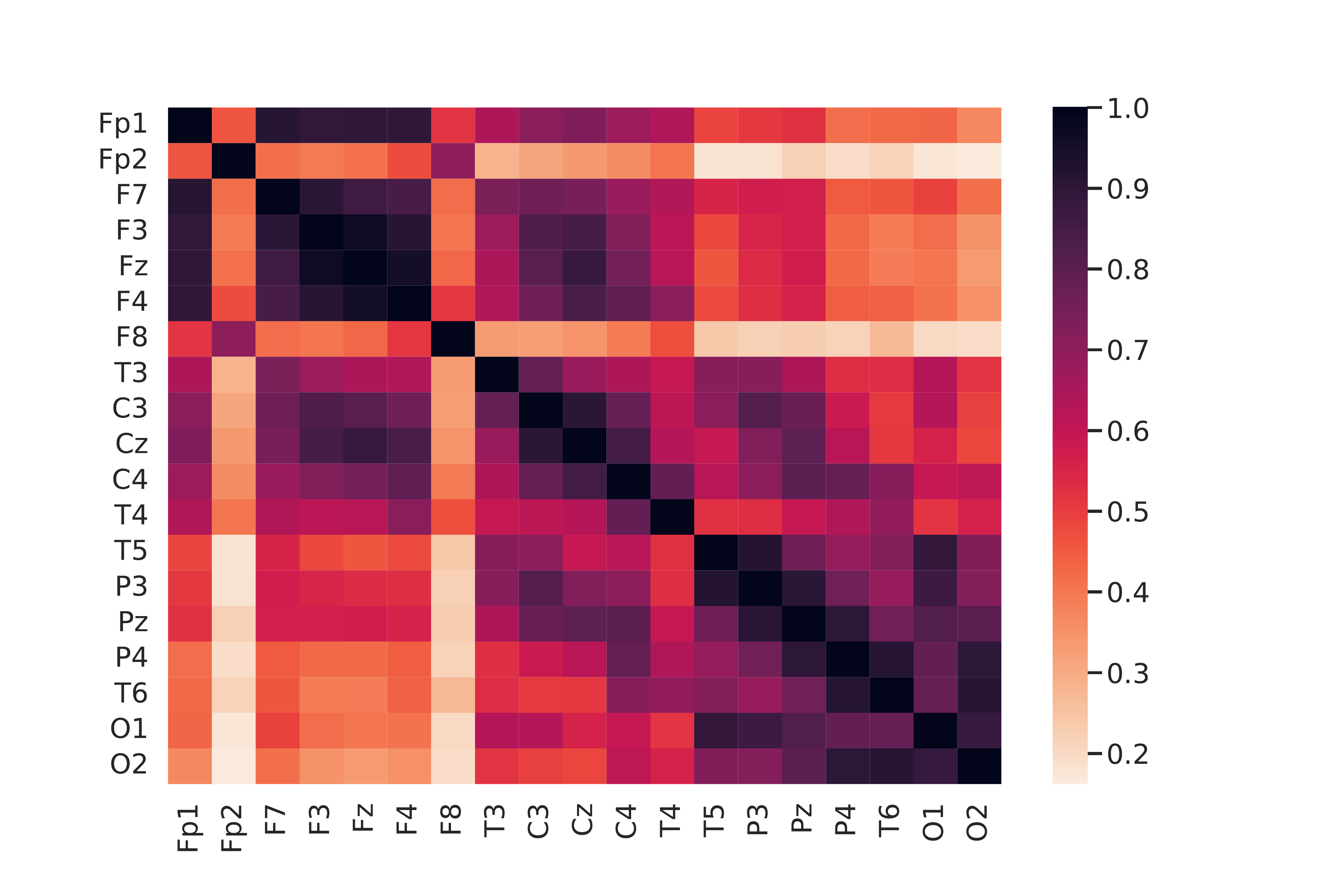}}}
\caption{Example of matrices of connections calculated with Pearson's correlation for (a) an individual with diagnosed AD and (b) an healthy individual.}
\label{Fig:matrices-pearson}
\end{figure*}

\begin{figure*}[!t]
\centering
\subfigure[]{
{\includegraphics[width=0.450\textwidth]{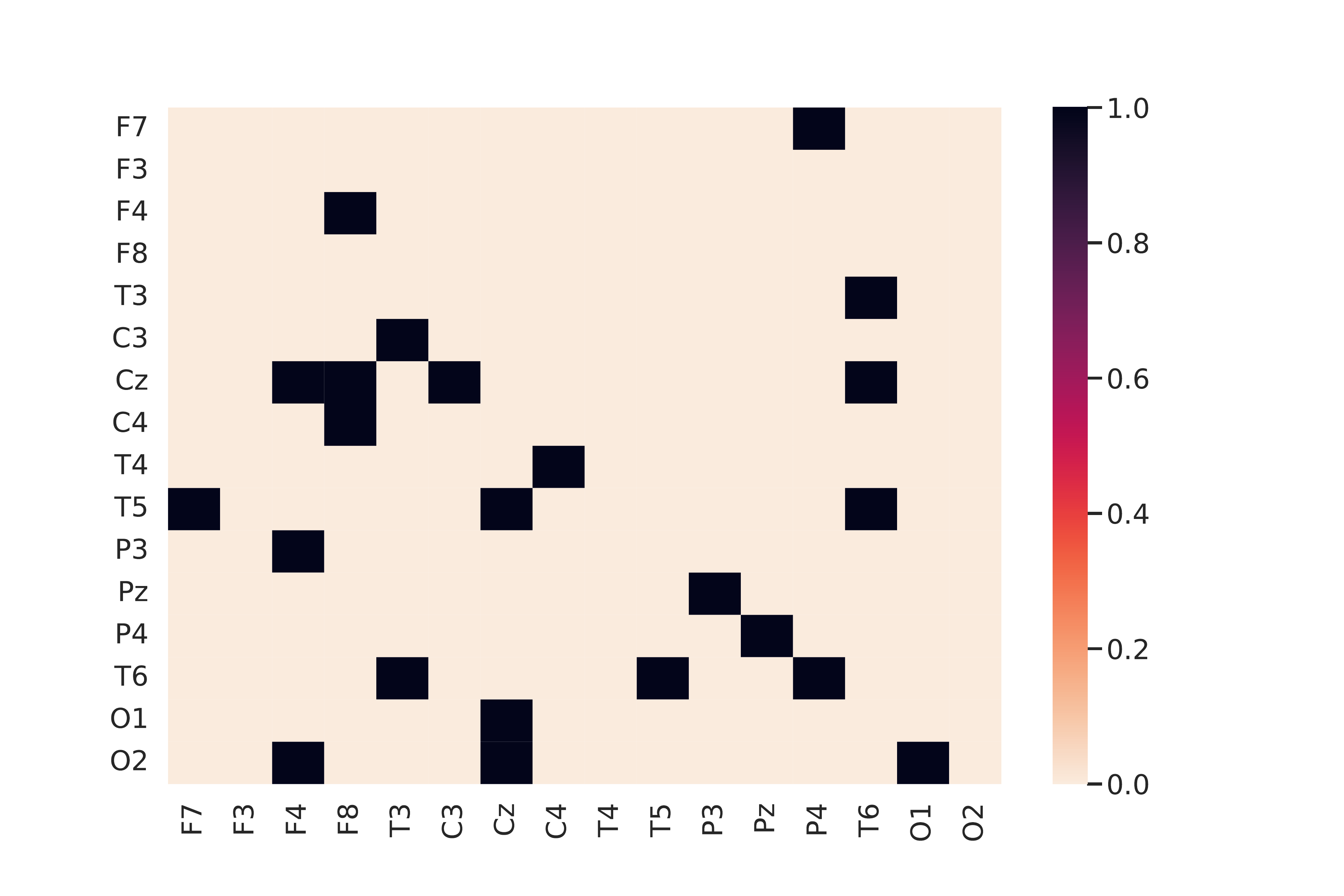}}}
\subfigure[]{
{\includegraphics[width=0.450\textwidth]{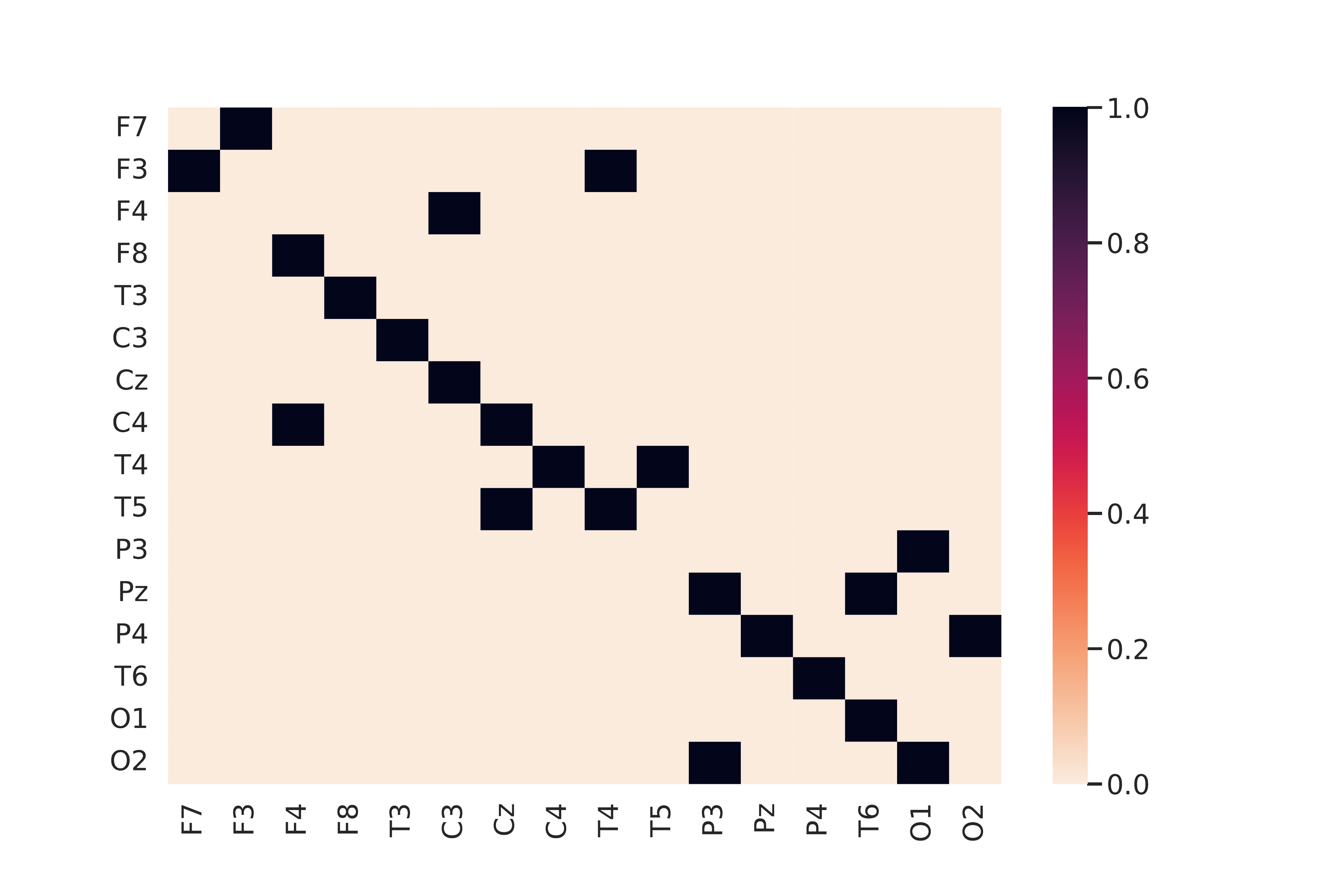} }}
\caption {Example of matrices of connections calculated with Granger causality test for (a) an individual with diagnosed SZ and (b) an healthy individual.}
\label{Fig:matrices-granger}
\end{figure*}

We consider the EEG time series described in section~\ref{Sec:database} to construct the matrices of connections for healthy controls and individuals diagnosed with AD and SZ, following the description in section~\ref{Sec:concepts}. These matrices are built by using the Granger causality test, Pearson's and Spearman's correlations measures for both data sets. In Figures \ref{Fig:matrices-pearson} and \ref{Fig:matrices-granger}, some examples of such matrices of connections are showed and differences between them can be noticed visually in both cases.  

The matrices of connections are inserted into the CNN by applying the flattening method, which converts the data into a 1-dimensional array that is input to the next layer.
Two CNN architectures are considered, i.e.
CNN$_{tuned}$  and CNN$_{untuned}$, to evaluate the classification. The CNN$_{tuned}$ is obtained by hyperparameter optimization, whereas the CNN$_{untuned}$ is a simpler model,  without using the tuning optimization. The evaluation of both models is done by using the area under the ROC curve (AUC). Nested k-fold cross-validation ($k=10$) for model selection, adjustment and evaluation is considered here.

The results for the  CNN$_{tuned}$ model is shown in tables \ref{Tab:ad} and~\ref{Tab:sz}, for AD and SZ, respectively. In all the cases, the CNN$_{tuned}$ model can unambiguously distinguish healthy individuals from individuals diagnosed with a brain disorder. The best results with an accuracy close to $100\%$ are obtained for both AD and SZ in the testing set using random search for hyperparameter tuning. 

\begin{table*} [!t]
\centering
\caption{Classification results for AD using the CNN$_{tuned}$ model (best results are in bold).}
\vspace{3mm}
\begin{tabular}{cccccccc}
\hline
\textbf{\begin{tabular}[c]{@{}c@{}} Matrices of\\ connections \end{tabular}}                  & \textbf{Hyperparameter}                                                                             & \textbf{Sample} & \textbf{Accuracy}  & \textbf{Precision}& \textbf{Recall} & \textbf{AUC}   \\ \hline
\multirow{6}{*}{\textbf{\begin{tabular}[c]{@{}c@{}}Granger\\ causality\end{tabular}}}    & \multirow{2}{*}{\textbf{\begin{tabular}[c]{@{}c@{}}Random\\ Search\end{tabular}}}         & \textbf{Train}  & 0.81           & 0.81             & 0.81         & 0.88              \\
                                                                                         &                                                                                            & \textbf{Test}   & 0.75            & 0.75             & 0.75          & 0.97              \\
                                                                                         & \multirow{2}{*}{\textbf{hyper-band}}                                                        & \textbf{Train}  & 0.65            & 0.65             & 0.65          & 0.65               \\
                                                                                         &                                                                                            & \textbf{Test}   & 0.75            & 0.75            & 0.75          & 0.97              \\
                                                                                         & \multirow{2}{*}{\textbf{\begin{tabular}[c]{@{}c@{}}Bayesian \\ Optimization\end{tabular}}} & \textbf{Train}  & 0.68            & 0.68             & 0.68          & 0.82               \\
                                                                                         &                                                                                            & \textbf{Test}   & 0.75            & 0.75             & 0.75          & 0.93              \\\hline
\multirow{6}{*}{\textbf{\begin{tabular}[c]{@{}c@{}}Pearson's \\ correlation\end{tabular}}} & \multirow{2}{*}{\textbf{\begin{tabular}[c]{@{}c@{}}Random\\ Search\end{tabular}}}          & \textbf{Train}  & \textbf{0.95}            & \textbf{0.95}             & \textbf{0.95}          & \textbf{0.98}               \\
                                                                                         &                                                                                            & \textbf{Test}   & \textbf{1.00}            & \textbf{1.00}             & \textbf{1.00}          & \textbf{1.00}           \\
                                                                                         & \multirow{2}{*}{\textbf{hyper-band}}                                                        & \textbf{Train}  & 0.86            & 0.86             & 0.86          & 0.90            \\
                                                                                         &                                                                                            & \textbf{Test}   & 1.00           & 1.00            & 1.00         & 1.00          \\
                                                                                         & \multirow{2}{*}{\textbf{\begin{tabular}[c]{@{}c@{}}Bayesian \\ Optimization\end{tabular}}} & \textbf{Train}  & 0.88            & 0.88            & 0.88         & 0.98               \\
                                                                                              &                                                                                      & \textbf{Test}   & 1.00            & 1.00             & 1.00          & 1.00          \\\hline
\multirow{6}{*}{\textbf{\begin{tabular}[c]{@{}c@{}}Spearman\\ correlation\end{tabular}}} & \multirow{2}{*}{\textbf{\begin{tabular}[c]{@{}c@{}}Random\\ Search\end{tabular}}}         & \textbf{Train} &0.47  & 0.47        & 0.45       & 0.47      \\
                                                                                         &                                                                                            & \textbf{Test}   & 0.75            & 0.75            & 0.75          & 0.75            \\
                                                                                         & \multirow{2}{*}{\textbf{hyper-band}}                                                        & \textbf{Train}  & 0.47            & 0.47             & 0.47          & 0.45              \\
                                                                                         &                                                                                            & \textbf{Test}   & 0.75           & 0.75            & 0.75          & 0.62               \\
                                                                                        & \multirow{2}{*}{\textbf{\begin{tabular}[c]{@{}c@{}}Bayesian \\ Optimization\end{tabular}}} & \textbf{Train}  & 0.47            & 0.47            & 0.47          & 0.45              \\
                                                                
                                                        &         &\textbf{Test}   & 0.75           & 0.75    & 0.75          & 0.68 \\ 
                                                                 \hline

\end{tabular}
\label{Tab:ad} 
\end{table*}

\begin{table*}
\centering
\caption{Classification results for SZ using the CNN$_{tuned}$ model (best results are in bold).}
\vspace{3mm}
\begin{tabular}{cccccccc}
\hline
\textbf{\begin{tabular}[c]{@{}c@{}} Matrices \\ of connections \end{tabular}}                  & \textbf{Hyperparameter}                                                                             & \textbf{Sample} & \textbf{Accuracy}  & \textbf{Precision}& \textbf{Recall} & \textbf{AUC}   \\ \hline
\multirow{6}{*}{\textbf{\begin{tabular}[c]{@{}c@{}}Granger\\ causality\end{tabular}}}    & \multirow{2}{*}{\textbf{\begin{tabular}[c]{@{}c@{}}Random\\ Search\end{tabular}}}          & \textbf{Train}   & \textbf{0.90}            & \textbf{0.90}            & \textbf{0.90}          & \textbf{0.93}           \\
                                                                                         &                                                                                            & \textbf{Test}              & \textbf{1.00}             & \textbf{1.00}         & \textbf{1.00}  & \textbf{1.00}             \\
                                                                                         & \multirow{2}{*}{\textbf{hyper-band}}                                                        & \textbf{Train}  & 0.73           & 0.73             & 0.73          & 0.77           \\
                                                                                         &                                                                                            & \textbf{Test} & 0.72            & 0.72             & 0.72          & 0.78            
  \\
                                                                                         & \multirow{2}{*}{\textbf{\begin{tabular}[c]{@{}c@{}}Bayesian \\ Optimization\end{tabular}}} & \textbf{Train} & 0.72            & 0.72             & 0.72          & 0.78            \\
                                                                                         &                                                                                           & \textbf{Test}   & 1.00            & 1.00             & 1.00          & 1.00            
  \\\hline
\multirow{6}{*}{\textbf{\begin{tabular}[c]{@{}c@{}}Pearson's \\ correlation\end{tabular}}} & \multirow{2}{*}{\textbf{\begin{tabular}[c]{@{}c@{}}Random\\ Search\end{tabular}}}     & \textbf{Train}  & 0.54            & 0.54             & 0.54          & 0.54            \\
                                                                                         &                                                                                            & \textbf{Test}   & 0.50            & 0.50             & 0.50          & 0.50              \\
                                                                                         & \multirow{2}{*}{\textbf{hyper-band}}                                                        & \textbf{Train}  & 0.54            & 0.54             & 0.54          & 0.54              \\
                                                                                         &                                                                                            & \textbf{Test}   & 0.50            & 0.50             & 0.50          & 0.50             \\
                                                                                         & \multirow{2}{*}{\textbf{\begin{tabular}[c]{@{}c@{}}Bayesian \\ Optimization\end{tabular}}} & \textbf{Train}  & 0.54            & 0.54             & 0.54          & 0.54             \\
                                                                                         &                                                                                            & \textbf{Test}   & 0.50            & 0.50             & 0.50          & 0.50            \\\hline

\multirow{6}{*}{\textbf{\begin{tabular}[c]{@{}c@{}}Spearman\\ correlation\end{tabular}}} & \multirow{2}{*}{\textbf{\begin{tabular}[c]{@{}c@{}}Random\\ Search\end{tabular}}}          & \textbf{Train}  & 0.53            & 0.53             & 0.53          & 0.53            \\
                                                                                 &                                                                                            & \textbf{Test}   & 0.50            & 0.50          & 0.50          & 0.50            \\
                                                                                         & \multirow{2}{*}{\textbf{hyper-band}}                                                        & \textbf{Train}  & 0.53            & 0.53             & 0.53          & 0.53             \\
                                                                                         &                                                                                            & \textbf{Test}   & 0.50            & 0.50             & 0.50          & 0.50              \\
                                                                                       & \multirow{2}{*}{\textbf{\begin{tabular}[c]{@{}c@{}}Bayesian \\ Optimization\end{tabular}}} & \textbf{Train}  & 0.53            & 0.53             & 0.53          & 0.53             \\
                                                                                         &                                                                                            & \textbf{Test}   & 0.50            & 0.50             & 0.50          & 0.50     
                                                                                 \\ \hline

\end{tabular}
\label{Tab:sz}
\end{table*}

Concerning the CNN$_{untuned}$ model, the results are shown in tables \ref{Table:cass-ad} and \ref{Table:class-sz} for AD and SZ, respectively.  For the AD data set, the best results are found using Pearson's correlation with a test accuracy of $92\%$. Regarding SZ disease, independently of the method used for the construction of the matrices of connections,  results are close to the random guessing  (see  table \ref{Table:class-sz}). Therefore, the CNN$_{tuned}$ model is more accurate for both AD and SZ diagnosis.

\begin{table*}[!t] 
\centering
\caption{Classification results for AD using the CNN$_{untuned}$ model (best results are in bold).}
\begin{tabular}{ccccccl}
\hline
\textbf{Matrices of connections}                  & \textbf{Sample} & \textbf{Accuracy} & \textbf{Precision} & \textbf{Recall} & \multicolumn{1}{l}{\textbf{AUC}} \\ \hline
\multirow{2}{*}{\textbf{Granger causality}}  & \textbf{Train}  & 0.97              & 0.97               & 0.99           & 0.99 \\
                                   & \textbf{Test}   & 0.58              & 0.57              & 0.66          & 0.75                              \\
\multirow{2}{*}{\textbf{Pearson's correlation}}  & \textbf{Train}  & \textbf{0.98}           & \textbf{0.99}               & \textbf{0.98}            & \textbf{0.99}                                                 \\
                                   & \textbf{Test}   & \textbf{0.92}           & \textbf{1.00}               & \textbf{0.83}          & \textbf{1.00}       \\
\multirow{2}{*}{\textbf{Spearman's correlation}} & \textbf{Train}  & 0.97             & 0.98               & 0.97            & 0.99                                              \\
                                   & \textbf{Test}   & 0.83              & 1.00               & 0.66           & 1.00                                               \\ \hline
\end{tabular} \label{Table:cass-ad}
\end{table*}

\begin{table*}[!t]
\centering
\caption{Classification results for SZ using the CNN$_{untuned}$ model.}
\begin{tabular}{ccccccl}
\hline
\textbf{Matrices of connections}                  & \textbf{Sample} & \textbf{Accuracy} & \textbf{Precision} & \textbf{Recall} & \multicolumn{1}{l}{\textbf{AUC}} \\ \hline
\multirow{2}{*}{\textbf{Granger causality}}  & \textbf{Train}  & \textbf{0.97}              & \textbf{0.97}               & \textbf{0.97}            & \textbf{0.99} \\
                                   & \textbf{Test}   & \textbf{0.52}              & \textbf{0.53}               & \textbf{0.73}           & \textbf{0.55}                              \\
\multirow{2}{*}{\textbf{Pearson's correlation}}  & \textbf{Train}  & 0.61           & 0.58              & 1.00            & 0.53                                                \\
                                   & \textbf{Test}   & 0.57              & 0.55               & 1.00          & 0.45        \\
\multirow{2}{*}{\textbf{Spearman's correlation}} & \textbf{Train}  & 0.62             & 0.59               & 0.97           & 0.58                                              \\
                                   & \textbf{Test}   & 0.62             & 0.58               & 1.00          & 0.53                                               \\ \hline
\end{tabular} \label{Table:class-sz}
\end{table*}

Importantly, the overall predictive performance depends on the choice of measure to construct the matrices of connections. In the case of AD, Pearson's correlation provides the best performance in CNN$_{tuned}$ (see table \ref{Tab:ad}).  On the other hand, in the case of SZ, Granger causality is superior to the other methods (see table \ref{Tab:sz}).  Therefore, there is no general method to infer the connections and obtain the most accurate results. Thus, different methods should be considered to develop an accurate framework for the automatic diagnosis of mental disorders.

For a comparison of our method with the more common approach known from the literature, the classification is performed by applying the raw EEG time series as input for the CNN$_{tuned}$ model (whose performance is the best for both diseases, as discussed before). The results are shown in tables  \ref{Table:results-rawAlzheimer} and \ref{Table:results-rawschizophrenia} for AD and SZ, respectively. The accuracy of $75\%$ for AD and $55\%$ for SZ are obtained. This outcome is supported by results available in the literature. Janghel and Rathore~\cite{janghel2021deep} obtained an accuracy of $76\%$ for AD, where the authors did not consider the matrices of connections.

\begin{table}[!t]
\centering
\caption{
Classification results for AD using raw EEG time series and the CNN$_{tuned}$ model.}
\begin{tabular}{cccccc}
\hline
\textbf{Set}   & \textbf{Accuracy} & \textbf{Precision} & \textbf{Recall} & \textbf{AUC} \\ \hline
\textbf{Train} & 0.68            & 0.61            & 1.00          & 0.68            \\
\textbf{Test}  & 0.75           & 0.66             & 1.00          & 0.75          \\ \hline
\end{tabular}\label{Table:results-rawAlzheimer}

\end{table}

\begin{table}[!t]
\centering
\caption{Classification results for SZ using raw EEG time series and the CNN$_{tuned}$ model.}
\begin{tabular}{cccccc}
\hline
\textbf{Set}   & \textbf{Accuracy} & \textbf{Precision} & \textbf{Recall} & \textbf{AUC}  \\ \hline
\textbf{Train} & 0.62            & 0.62            & 1.00          & 0.50             \\
\textbf{Test}  & 0.55           & 0.55             & 1.00         & 0.50         \\ \hline
\end{tabular}\label{Table:results-rawschizophrenia}
\end{table}

\begin{figure*}
\centering
\subfigure[]{
\includegraphics[width=0.45\textwidth]{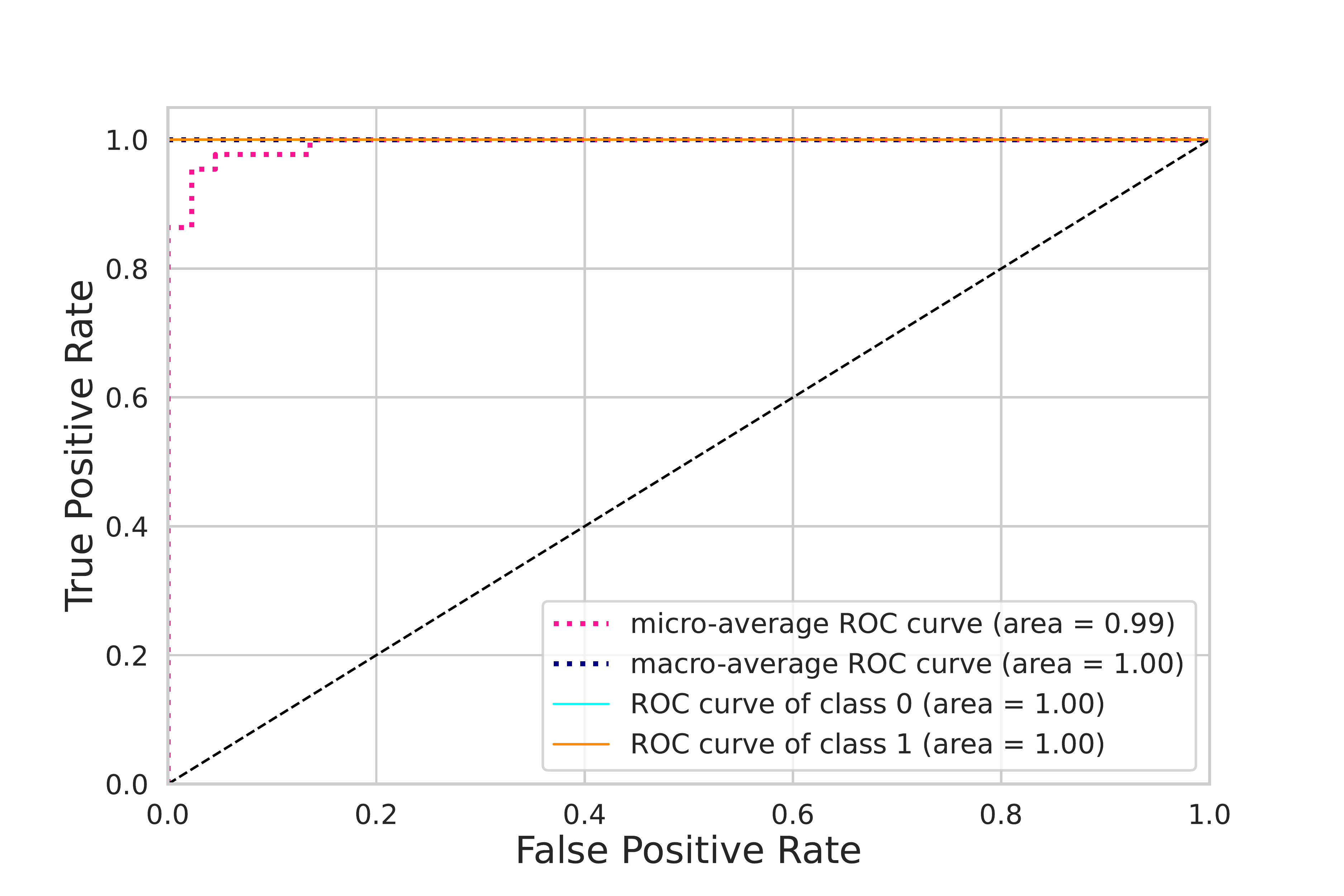}}
\subfigure[]{
\includegraphics[width=0.45\textwidth]{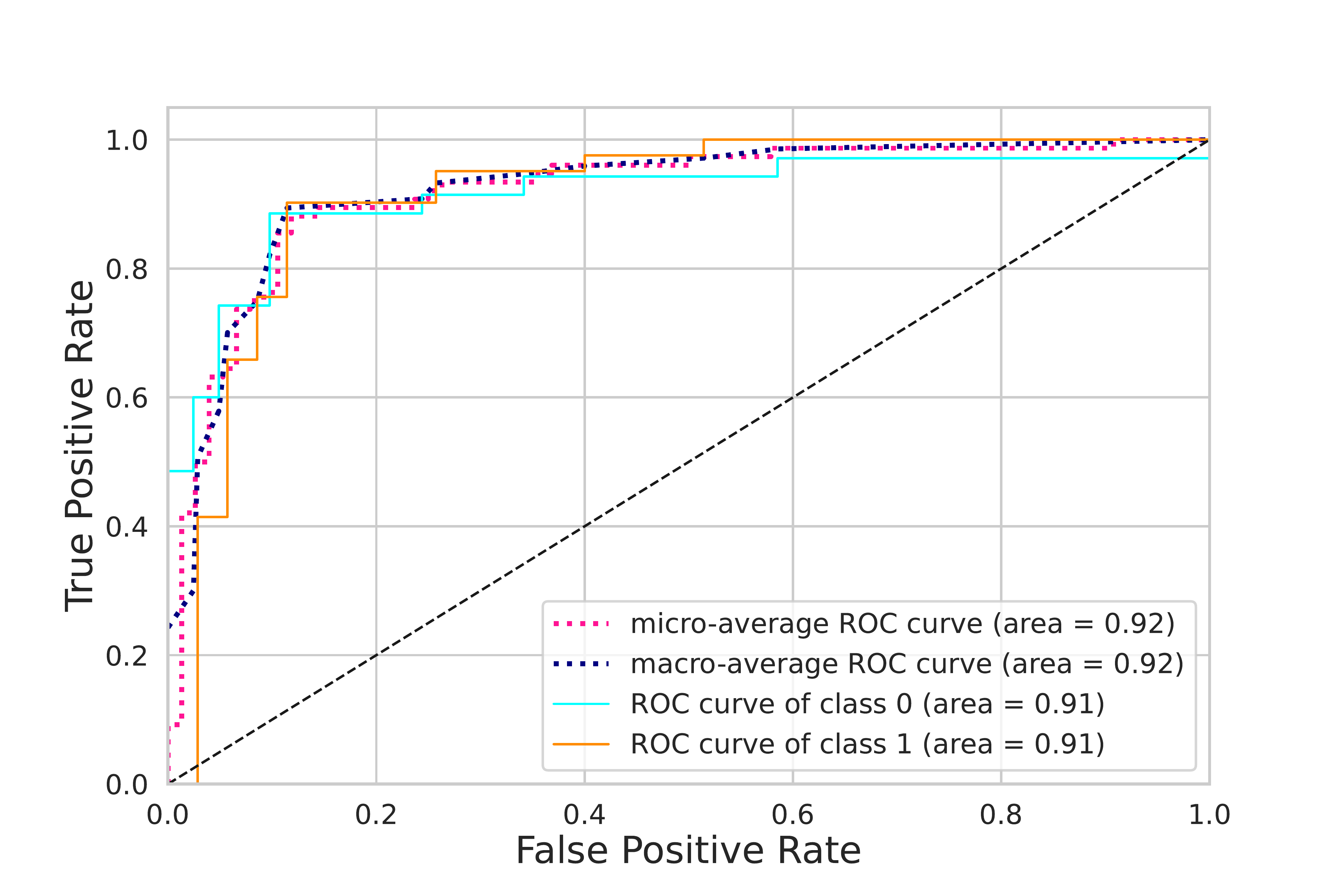}}
\caption{ROC curve obtained from the CNN$_{tuned}$ model. The matrices of connections are constructed by (a) Pearson's correlation for AD disease and (b) Granger causality for individuals diagnosed with SZ.}
\label{Fig:AUC-pearson-granger-schizophrenia-matrix}
\end{figure*}

As we can see, our proposed method based on a matrix of connections provided as input to a CNN allows for more accurate results. This reinforces the importance of using a data set that encompasses the connections between brain regions. Indeed, the network structure is a fundamental ingredient to differentiate healthy individuals from patients presenting neurological disorders, as verified in many papers (e.g.~\cite{de2014structure, lynn2019physics, fallani2011multiple, rodrigues2009structure, antiqueira2010estimating}).

\begin{figure*}
\centering
\subfigure[]{
\includegraphics[width=0.45\textwidth]{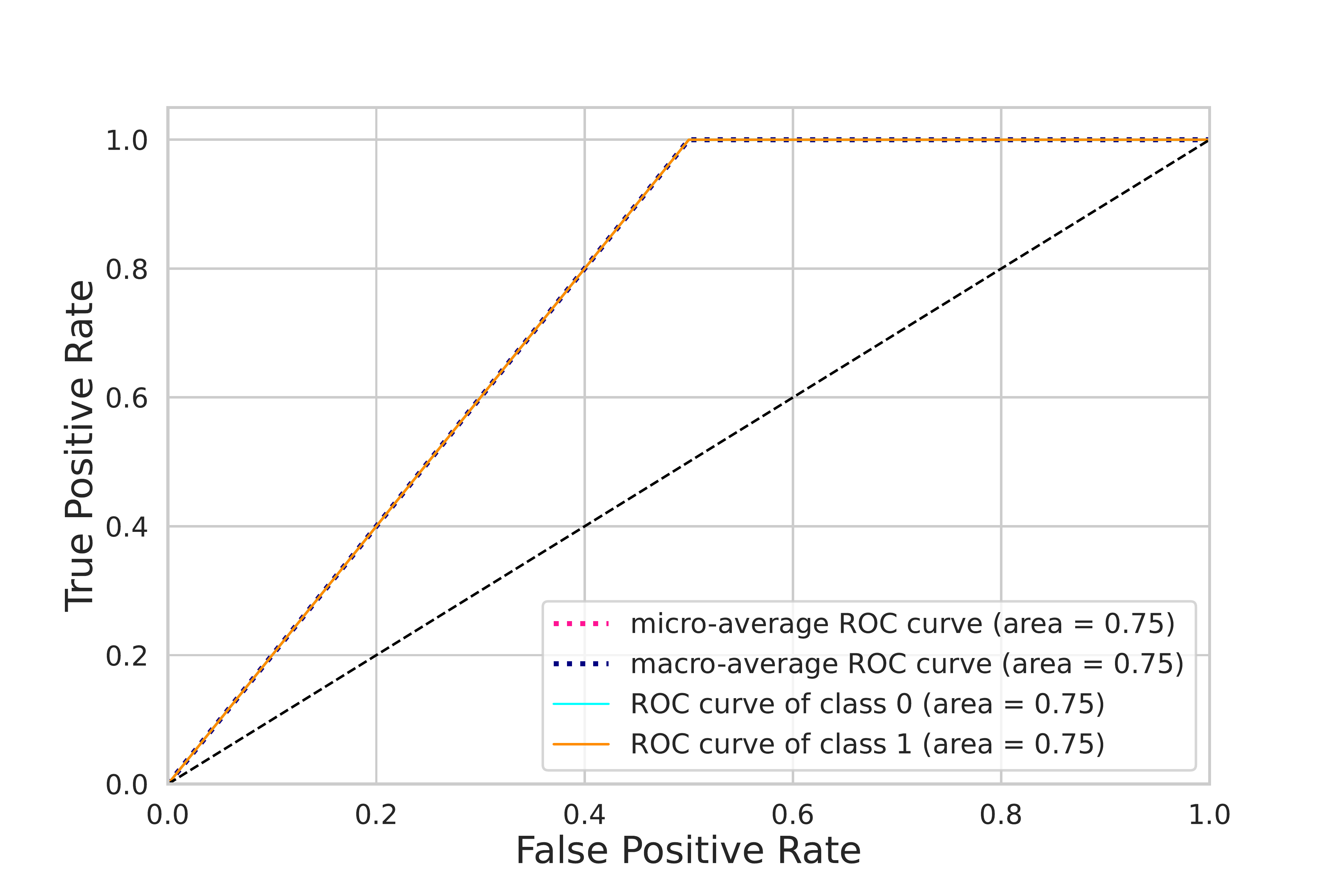}}
\subfigure[]{
\includegraphics[width=0.45\textwidth]{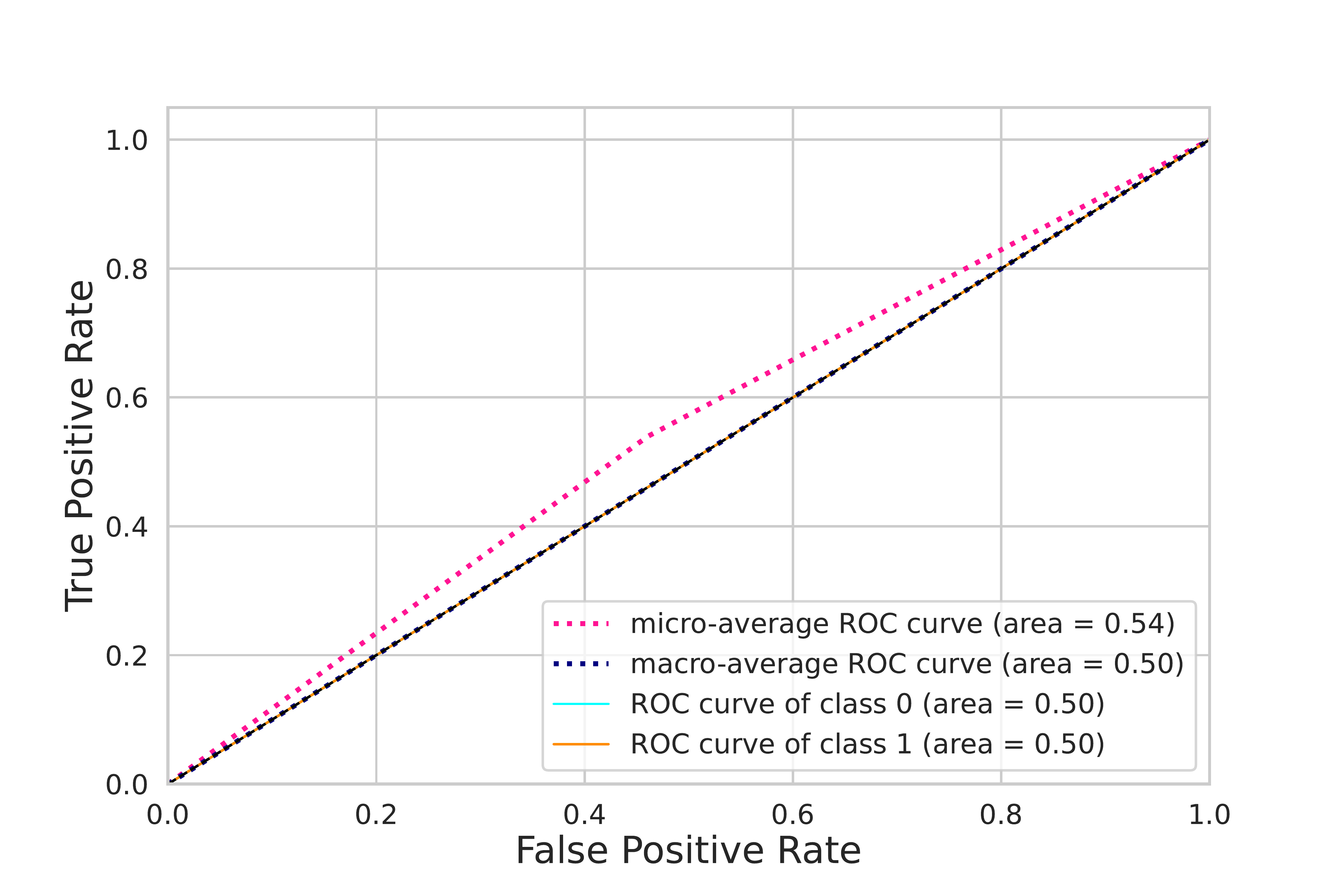}}
\caption{ROC curve obtained from raw EEG time series for (a) individuals diagnosed with AD and (b) individuals with SZ.}
\label{Fig:AUC-pearson-granger-schizophrenia}
\end{figure*}

In Figures \ref{Fig:AUC-pearson-granger-schizophrenia-matrix} we show the ROC curve for the best results, i.e. for AD (using Pearson's correlation) and SZ (using Granger causality test), respectively. For AD, the micro and macro-average ROC curve areas are 0.99 and 1.0, respectively, the micro and macro-average ROC curve areas are 0.92 for both cases. For comparison, Figure \ref{Fig:AUC-pearson-granger-schizophrenia} shows the ROC curve for AD and SZ using raw times series, where the micro and macro-average ROC curve areas are 0.75 for AD and around 0.55 for AZ. Comparing these results, we conclude that the use of the matrix of connections provides the most accurate classifications.

\section{Conclusion}

In this paper, we propose a method for automatic diagnosis of AD and SZ based on EEG time series and deep learning. We infer the matrix of connections between brain areas following three different approaches, based on Granger causality, Pearson's and Spearman's correlations. These matrices are included in a convolutional neural network, tunned with the random search, hyper-band, and Bayesian optimization. We verify that this approach provides a very accurate classification of patients with AD and SZ diseases. The comparison with the traditional method that considers raw EEG data shows that our method is more accurate, reinforcing the importance of network topology for the description of brain data. Our method is general and can be used for any mental disorder in which EEG times series can be recorded. 

A limitation of our analysis is the relatively small data set, although this is common in other studies on disease classification \cite{oh2019deep}. However, even with this restriction, our algorithm worked very well, showing that AD and SZ are associate to changes in brain organization. As future work, we suggest to consider larger data sets and additional information about the patients, like health conditions and age. A method that provides the level of the evolution of the disease is also an interesting topic to be developed from our study.

\section{Acknowledgements}

F.A.R. acknowledges CNPq (grant 309266/2019- 0) and FAPESP (grant 19/23293-0) for the financial support given for this research. A.M.P. acknowlwdges FAPESP (grant 2019/22277-0) for the financial support given this research.
K.R. acknowledges FAPESP grant 2019/26595-7.
C.T.  gratefully acknowledges financial support
from the Zentrum für Wisschenschaftliche Services und
Transfer (ZeWiS) Aschaffenburg, Germany.

\bibliography{references}

\end{document}